\documentclass[12pt,a4paper]{article}
\usepackage{jheppub}

\title{Superluminal Vector in Ghost-free Massive Gravity}
\author[a]{Siqing Yu}
\affiliation[a]{Department of Physics, Columbia University,\\
538 West 120th Street, New York, NY 10027, USA}    
\emailAdd{sy2384@columbia.edu}
\abstract{We present a classical analysis on the issue of vector superluminality in the decoupling limit ghost-free massive gravity with a Minkowski reference metric. We show explicitly in the Lorenz gauge that the theory is free of superluminal vector excitations around a nontrivial solution at the cubic order in the fields. In the same gauge, we demonstrate that superluminal vector modes arise at the quartic order and compute some superluminal propagating solutions. We then generalize our findings to all orders in a gauge-independent way. We check the physical consistency of the vector superluminalities, arguing that they are not physically detectable in the perturbation theory but could be trusted classically in the strong coupling region. Nevertheless, these superluminalities involve only low frequency group and phase velocities and are unable to determine the acausality of the theory.}

\arxivnumber{1310.6469}

\usepackage{breqn}
\makeatletter
\def\@fpheader{~~~~~~~~~~~~~~~~~~~~~~~~~~~~~~~}
\makeatother

\begin{document}
\maketitle
\numberwithin{equation}{section}

\section{Introduction and Summary}
Einstein's General Relativity (GR) is a simple, elegant and successful theory of the intrinsic spacetime structure of the universe. Since its inauguration in 1916, it has been verified by many experimental tests and found itself accountable for numerous technological advances. Yet the recent discovery of the accelerated cosmic expansion \cite{Perlmutter:1997zf,Riess:1998cb,Tonry:2003zg} and the need to address the cosmological constant problem \cite{Weinberg1} sparked renewed interest in the search for local infrared modifications of GR. Among these theoretical attempts, one possibility is to give the graviton, the physical construction that is supposed to mediate interactions in gravity, a small mass of order of the contemporary value of the Hubble constant. This is the so-called massive gravity.\footnote{For reviews on massive gravity, see Refs. \cite{Hinterbichler:2011tt,deRham:2014zqa}.}

The unique simplest linear theory of massive gravity was proposed by Fierz and Pauli in 1939 \cite{Fierz:1939ix}. It describes to the linear order a single massive spin-2 particle with the consistent mode functions (5 degrees of freedom: two helicity-2, two helicity-1 and one helicity-0) in the high energy limit. In the past, various curiosities were raised about the Fierz-Pauli (FP) theory, such as the van Dam-Veltman-Zakharov (vDVZ) discontinuity \cite{vanDam:1970vg,Zakharov:1970cc}, which prevents the theory from reducing smoothly to GR in the massless limit, and the Boulware-Deser ghost \cite{Boulware:1973my}, which plagues the theory with a sixth degree of freedom. Hearteningly, in the recent years a non-linear theory correcting FP with a 2-parameter family of higher order potential terms has been proposed \cite{deRham:2010kj,deRham:2010ik}. This non-linear theory, first formulated by de Rham, Gabadadze and Tolley (hence commonly known as dRGT), capitalizes on the Vainshtein mechanism \cite{Vainshtein:1972sx}, one that screens the effect of fifth forces with the non-linearities of the helicity-0 mode, as a cure to the vDVZ discontinuity. The theory is also proved to be free of Boulware-Deser ghost around any background metric in a myriad of languages and formalisms describing the theory \cite{Hassan:2011hr,Hassan:2011ea,deRham:2011rn,deRham:2011qq,Mirbabayi:2011aa,Deffayet:2012nr,Hinterbichler:2012cn}.

However, there remains the important point of contention about the ghost-free dRGT massive gravity, which is whether the theory is free of superluminal fluctuations around nontrivial background solutions. In the decoupling limit,\footnote{See (\ref{eq:8}).} dRGT eliminates the Boulware-Deser ghost by introducing Galileon scalar self-interaction terms for the helicity-0 mode \cite{Nicolis:2008in,deRham:2010ik}. In general, these Galileon terms may lead to solutions of the dRGT field equations where the scalar excitations exhibit superluminal group and phase velocities, starting from the cubic order terms onwards \cite{Nicolis:2008in,Nicolis:2009qm,deFromont:2013iwa,Adams:2006sv}. This feature is present outside the decoupling limit as well \cite{Deser:2012qx,Izumi:2013poa,Deser:2013eua}. Yet many such solutions are shown to be unstable \cite{Gruzinov:2011sq,deRham:2011pt,Berezhiani:2013dw,Berezhiani:2013dca}. Also, it can be argued that the superluminality in the scalar sector of Galileon theories and massive gravity does not meddle with the usual notion of causality in local Lorentz invariant theories \cite{deFromont:2013iwa,Burrage:2011cr}. Meanwhile, as a side note, the freedom to choose the values of the two free parameters in dRGT, named $c_3$ and $d_5$ as they appear in the metric formulation \cite{deRham:2010ik}, would circumvent the issue of scalar superluminality in the decoupling limit for a special choice of the parameters.\footnote{In the vierbein formalism, which is dynamically equivalent to the metric formulation, the couplings are related to $c_3$ and $d_5$ as follows, up to an irrelevant overall constant: $\beta_{1}=6-36c_3-96d_5$, $\beta_2=-2+24c_3+96d_5$, $\beta_3=-12c_3-96d_5.$} By setting $c_{3}=1/6$ and $d_{5}=-1/48$, the scalar self-interaction and the terms mixing spin-2 and -0 fields are completely eliminated from the dRGT Lagrangian \cite{deRham:2010ik,Hinterbichler:2011tt}. The resulting ``minimal model" is then that of free helicity-2 and -0 modes. 

With these technicalities extensively discussed in the previous literature, a natural step is then to investigate superluminality in the vector sector of the dRGT massive gravity, which we hope to cover as thoroughly as possible in this paper.

Our work is organized as follows. In section 2, we review the metric formulation of dRGT theory and calculate the St\"{u}kelberg vector Lagrangian to the cubic order in Minkowski background metric, keeping $c_{3}$ explicitly as a free parameter in the process. At the level of equation of motion, we show that the vector fluctuations cannot be superluminal. In section 3, we go to the quartic order and derive the effective equations of motion with the dangerous term, suppressing all other terms. We show that superluminal propagating solutions na\"{i}vely exist at this order by constructing an example on a representative background solution and characterizing the superluminal modes. In section 4, we proceed in a gauge-independent fashion and analyze vector superluminality to all orders, reinforcing and generalizing our results in section 3 by a degree of freedom count in the Hamiltonian formalism. We then check the consistency of the superluminalities. We find that in the linear regime where perturbativity holds, superluminal vector excitations are not physically observable, whereas in the non-linear strong coupling region, superluminal signals could possibly arise and be detected. We discuss the implications of our findings to the dRGT massive gravity as well as suggest possible future work in section 5.
\section{Cubic Order}
\subsection{dRGT Ghost-free Potential}
The dRGT Lagrangian for a massive spin-2 field \cite{deRham:2010ik} is given by
\begin{equation} \label{eq:1}
\mathcal{L}=M_{\text{P}}^2\sqrt{-g}R-\frac{M_{\text{P}}^2m^2}{4}\sqrt{-g}\left( V_{2}(g, H)+V_{3}(g, H)+V_{4}(g, H)+V_{5}(g, H)+...\right),
\end{equation}
where $M_{\text{P}}$ is the Planck mass, $m$ the graviton mass, and $V_{i}$ gives the interaction terms at $i$\textsuperscript{th} order in $H_{\mu \nu}$,
\begin{equation} \label{eq:2}
V_{2}(g, H)=[H^2]-[H]^2,
\end{equation}
\begin{equation} \label{eq:3}
V_{3}(g, H)=c_{1}[H^3]+c_{2}[H][H^2]+c_{3}[H]^3,
\end{equation}
\begin{equation} \label{eq:4}
V_{4}(g, H)=d_{1}[H^4]+d_{2}[H][H^3]+d_{3}[H^2]^2+d_{4}[H]^2[H^2]+d_{5}[H]^4,
\end{equation}
\begin{dmath} \label{eq:5}
V_{5}(g, H)=f_{1}[H^5]+f_{2}[H][H^4]+f_{3}[H]^2[H^3]+f_{4}[H^2][H^3]+f_{5}[H][H^2]^2+f_{6}[H]^3[H^2]+f_{7}[H]^5,
\end{dmath}
where the indices are contracted with the inverse metric, so that $[H]=g^{\mu \nu}H_{\mu \nu}$, $[H^2]=g^{\mu \nu}g^{\alpha \beta}H_{\mu \alpha}H_{\nu \beta}$, etc. The coefficients $c_{i}, d_{i}, f_{i}$ are related such that no ghosts exist up to the quintic order in the decoupling limit. By convention, $c_{3}, d_{5}$ and $f_{7}$ are chosen to carry the free parameters in the theory. In particular, 
\begin{equation} \label{eq:6}
c_{1}=2c_{3}+\frac{1}{2} \text{  and  } c_{2}=-3c_{3}-\frac{1}{2}.
\end{equation}

By requiring that the background metric $g_{\mu \nu}^{(0)}$ is general coordinate covariant, one can obtain the St\"ukelberg expansion for $H_{\mu \nu}(x)=g_{\mu \nu}(x)-g_{\alpha \beta}^{(0)}(Y(x))\partial_{\mu}Y^{\alpha}\partial_{\nu}Y^{\beta}$ \cite{ArkaniHamed:2002sp,Schwartz:2003vj}, where $Y^{\alpha}(x)$ are the four fields that transform as scalars under diffeomorphisms \cite{Hinterbichler:2011tt}: 
\begin{dmath} \label{eq:7}
H_{\mu \nu}=h_{\mu \nu}+\partial_{\mu}A_{\nu}+\partial_{\nu}A_{\mu}+2\partial_{\mu}\partial_{\nu}
\phi-\partial_{\mu}A^{\alpha}\partial_{\nu}A_{\alpha}-\partial_{\mu}A^{\alpha}\partial_{\nu}\partial_{\alpha}\phi
-\partial_{\nu}A^{\alpha}\partial_{\mu}\partial_{\alpha}\phi
-\partial_{\mu}\partial^{\alpha}\phi\partial_{\nu}\partial_{\alpha}\phi,
\end{dmath}
where we expand around a flat metric. From here onwards, the indices are raised with $\eta^{\mu \nu}$. The full dRGT Lagrangian in St\"ukelberg fields can then be derived by substituting (\ref{eq:6}) and (\ref{eq:7}) into (\ref{eq:1}). The terms with helicity-2 and -0 modes in the decoupling limit,
\begin{equation} \label{eq:8}
m\rightarrow 0, \text{ } M_{\text{P}}\rightarrow \infty, \text{ } \Lambda_{3}\equiv (m^2M_{\text{P}})^{1/3} \text{ and } \frac{T^{\mu \nu}}{M_{\text{P}}} \text{   fixed }
\end{equation}
have been worked out in Ref. \cite{deRham:2010ik}.

Here we need to emphasize that $\Lambda_3$ is called the strong coupling scale of the dRGT massive gravity; it signifies the scale at which perturbativity breaks down and and tree-level calculations no longer give the full picture. However, $\Lambda_3$ is {\it not} the energy cutoff of the theory -- it does not necessarily mean the beginning of new physics because quantum corrections are still suppressed by powers of the Planck scale, and we need the scalar St\"ukelberg field to take large values in a region where $\phi \sim \Lambda_3, \partial\phi \sim \Lambda_3^2, \partial^2\phi \sim \Lambda_3^3, \partial^n\phi \ll \Lambda_3^{n+1}$ for $n \geq 3$ for the Vainshtein mechanism to take effect \cite{Chkareuli:2011te,Nicolis:2004qq,Hinterbichler:2011tt}. Presumably, the cutoff could be below the Planck scale but much higher than the (redressed) strong coupling scale \cite{deRham:2014zqa}. It is crucial to recognize this in our discussion later on the physicality of superluminal modes.

\subsection{dRGT Vector Lagrangian up to the Cubic Order}
Since the vector modes do not appear linearly in the full dRGT Lagrangian, they can be savely set to zero in a trivial solution. However, vector fluctuations can still emerge around nontrivial solutions and they might be superluminal. As a preliminary attempt to clarify this, the vector Lagrangian in the decoupling limit up to the cubic order is calculated,\footnote{A maximally symmetric special case of this is given in Ref. \cite{Tasinato:2012ze}.}
\begin{dmath} \label{eq:2.9}
\mathcal{L}_{A}=-\frac{1}{4}F^{\mu \nu}F_{\mu \nu}-\frac{1}{4\Lambda_{3}^3}\partial^{\mu}\partial^{\nu}\phi \bigg[ (24c_{3}+2)\partial_{\mu}A^{\alpha}\partial_{\alpha}A_{\nu}
+(12c_{3}-1)\partial_{\mu}A^{\alpha}\partial_{\nu}A_{\alpha}+(12c_{3}-1)\partial^{\alpha}A_{\mu}\partial_{\alpha}A_{\nu}
-48c_{3}\partial_{\mu}A_{\nu}\partial^{\alpha}A_{\alpha} \bigg]
\\
-\frac{1}{4\Lambda_{3}^3}\Box\phi \bigg[ -(12c_{3}+2)\partial^{\mu}A^{\nu}\partial_{\nu}A_{\mu}
-(12c_{3}-2)(\partial_{\mu}A_{\nu})^2+24c_{3}(\partial^{\alpha}A_{\alpha})^2
\bigg],
\end{dmath}
after performing the canonical renormalization (with the convention in Ref. \cite{deRham:2010ik}) given by
\begin{equation} \label{eq:2.10}
A_{\mu}\rightarrow\frac{1}{mM_{\text{P}}}A_{\mu}  \text{     and  }  \phi \rightarrow \frac{1}{m^2M_{\text{P}}}\phi=\frac{1}{\Lambda_{3}^3}\phi.
\end{equation}
This gives the dRGT vector Lagrangian with arbitrary $c_{3}$ in the metric language. It is consistent with the result obtained in Ref. \cite{deRham:2010gu} Eqn. (22), where the model sets $c_{3}=1/4$. 
\subsection{Equations of Motion}
With some effort, the dRGT vector Lagrangian (\ref{eq:2.9}) can be rewritten as, up to a total derivative, 
\begin{dmath} \label{eq:2.11}
\mathcal{L}_{A}=-\frac{1}{4} \eta^{\alpha \beta} \left[ 
\eta^{\mu \nu} \left( 
1-\frac{6c_{3}-1}{\Lambda_{3}^3}\Box\phi \right) +\frac{12c_{3}-1}{\Lambda_{3}^3}\partial^{\mu}\partial^{\nu}\phi
\right] F_{\mu \alpha}F_{\nu \beta},
\end{dmath}
where $F_{\mu \nu}=\partial_{\mu}A_{\nu}-\partial_{\nu}A_{\mu}$ is the usual Maxwell field strength tensor. Note that in this form, the vector Lagrangian is invariant under the change of indices $\mu \nu \longleftrightarrow \alpha \beta$ and the U(1) gauge transformation $A_{\mu} \rightarrow A_{\mu}+\partial_{\mu} \pi$. Because (\ref{eq:2.11}) is quadratic in the vector field, the Lagrangian for vector fluctuations around a nontrivial solution has the same form. For simplicity, we still denote these fluctuations with $A_{\mu}$ for the rest of this paper.

The equations of motion for $A_{\alpha}$,  $\partial_{\mu}\frac{\partial \mathcal{L}_{A}}{\partial(\partial_{\mu}A_{\alpha})}=\frac{\partial \mathcal{L}_{A}}{\partial A_{\alpha}}$, with the symmetries mentioned above, then read
\begin{equation} \label{eq:2.12}
\left( \eta^{\alpha \beta}+\frac{12c_{3}-1}{\Lambda_{3}^3}\partial^{\alpha}\partial^{\beta}\phi
\right)
\bigg[ \Box A_{\beta}-\partial_{\beta}(\partial^{\nu}A_{\nu})
\bigg]=0,
\end{equation}
where we have dropped the correction to the flat metric for simplicity as it does not affect the vector light cone \cite{Nicolis:2009qm}. We can then impose the Lorenz gauge $\partial^{\mu}A_{\mu}=0$ to set the divergence to zero.\footnote{We are allowed to do this because fixing the gauge does not interfere with superluminalities in the theory. Also, the result here is reaffirmed in a gauge-independent way in section 4.} As such, the equation of motion in Fourier space shows that the momenta of the vector fluctuations do not couple to the effective metric in front, which depends on some scalar solution $\phi=\phi_{0}(x)$, so their light cone remains the same as that in the flat metric. This proves that the dRGT massive gravity does not manifest superluminal vector fluctuations up to the cubic order, unlike the case with scalar fields \cite{deFromont:2013iwa}.

However, (\ref{eq:2.12}) is proportional to the Maxwell equations $\partial^{\nu}F_{\nu \beta}=0$, which imply that the vector excitations are exactly luminal at this order. It is therefore necessary to go to higher orders to see whether superluminal vector signals actually exist in dRGT.
\section{Quartic Order and the Rise of Superluminality}
\subsection{Quartic dRGT Vector Lagrangian}
The recent progresses in the vierbein formalism have worked out the decoupling limit dRGT action to all orders \cite{Gabadadze:2013ria,Ondo:2013wka}, where it is clear that the {\it full} action depends on the helicity-1 mode only through terms manifestly quadratic in $F_{\mu \nu}$. By collecting relevant terms from Eqn. (3.36) of Ref. \cite{Ondo:2013wka} and making proper changes in the field conventions, we are able to reproduce (\ref{eq:2.11}). However, beginning with the quartic order, it will prove beneficial to rewrite the vector Lagrangian in a more illuminating form before working out the relevant terms with brute force.

In the decoupling limit, the helicity-1 terms are in the form $\sim(\partial^2\phi)^n(\partial A)^2$ \cite{Hinterbichler:2011tt,Koyama:2011wx}, so the vector Lagrangian looks like
\begin{equation} \label{eq:3.1}
\mathcal{L}_{A}=-\frac{1}{4} \mathcal{T}^{\mu \nu \alpha \beta} F_{\mu \alpha}F_{\nu \beta},
\end{equation}
where the 4-tensor $\mathcal{T}^{\mu \nu \alpha \beta}$ is constructed with only $\eta^{\mu \nu}$ and powers of $\Pi^{\mu \nu}\equiv \partial^\mu \partial^\nu \phi$ in a perturbative expansion. Here the St\"ukelberg fields are those after performing the canonical renormalization in (\ref{eq:2.10}). The full effective tensor then must take the form
\begin{equation} \label{eq:3.2}
\mathcal{T}^{\mu \nu \alpha \beta}=\sum_{p,q,\vec{l}}C_{p,q,\vec{l}=(l_1,l_2,...,l_N)}\underbrace{[\Pi]^{l_1}[\Pi^2]^{l_2}\cdots[\Pi^N]^{l_N}}_{\equiv F_{\vec{l}}(\Pi)}(\Pi^{\mu \nu})^{p}(\Pi^{\alpha \beta})^{q}.
\end{equation}
Here the notation needs some elaboration. First, $p$ and $q$ are summed over all natural numbers including zero, $\mathbb{N}\equiv \{ 0 \} \cup \mathbb{Z}_+$, with $p\leq q$, and we define $(\Pi^{\mu \nu})^0\equiv \eta^{\mu \nu}$. Second, $\vec{l}$ is summed over all finite-length natural number tuples whose last entries are nonzero (with the only exception $\vec{l}=(0)$, whose meaning is obvious). For clarity, this is the set
\begin{equation} \label{eq:3.3}
\{ \vec{l} \}=\{ (0) \} \cup \bigcup_{n\in \mathbb{N}}(\mathbb{N}^n \times \mathbb{Z}_+),
\end{equation}
whose countability follows from the theorems regarding countable sets \cite{Munkres}. Thus we can relabel the $\vec{l}$'s with natural numbers $r\in \mathbb{N}$,
\begin{equation} \label{eq:3.4}
\mathcal{T}^{\mu \nu \alpha \beta}=\sum_{p,q,r}C_{pqr}{F_{r}(\Pi)}(\Pi^{\mu \nu})^{p}(\Pi^{\alpha \beta})^{q},
\end{equation}
where $C_{pqr}$ are numerical constants and $F_r(\Pi)$ are scalar quantities. The specific relabeling scheme does not concern us here, but with some intuition we can let $F_0(\Pi)=1$. With the tensor given in (\ref{eq:3.4}), (\ref{eq:3.1}) already represents the most general form of the vector Lagrangian because of the identities
\begin{equation} \label{eq:3.5}
(\Pi^{\mu \beta})^{p}(\Pi^{\nu \alpha})^{q}F_{\mu \alpha}F_{\nu \beta}=-(\Pi^{\mu \nu})^{p}(\Pi^{\alpha \beta})^{q}F_{\mu \alpha}F_{\nu \beta}
\end{equation}
and
\begin{equation} \label{eq:3.6}
(\Pi^{\mu \nu})^{q}(\Pi^{\alpha \beta})^{p}F_{\mu \alpha}F_{\nu \beta}=(\Pi^{\mu \nu})^{p}(\Pi^{\alpha \beta})^{q}F_{\mu \alpha}F_{\nu \beta},
\end{equation}
which follow from the symmetry of $\Pi^{\mu \nu}$, the antisymmetry of $F_{\mu \nu}$ and the $\mu \nu \longleftrightarrow \alpha \beta$ symmetry of $F_{\mu \alpha}F_{\nu \beta}$.

Next, we make an important observation. With the restriction $p\leq q$, any term in the expansion (\ref{eq:3.4}) must have $p=0$ if it contains a factor of $\eta^{\mu \nu}$ or $\eta^{\alpha \beta}.$ Such terms do not lead to superluminality because their contribution to the equation of motion of $A_{\alpha}$ is
\begin{equation} \label{eq:3.7}
\sim \frac{\partial ( F_r(\Pi)\eta^{\mu \nu}(\Pi^{\alpha \beta})^qF_{\mu \alpha}F_{\nu \beta})}{\partial(\partial_{\mu}A_{\alpha})} \sim F_r(\Pi)\eta^{\mu \nu}(\Pi^{\alpha \beta})^q\partial_\mu F_{\nu \beta} \propto \partial^{\nu}F_{\nu \beta},
\end{equation}
which is merely a correction to the Maxwell term. This sufficiently explains why superluminality does not occur at the cubic order, as all the cubic order terms are characterized by $(p,q)=(0,0)$ or $(0,1)$.

At the quartic order, we have the following possibilities for $(p,q):(0,0),(0,1),(0,2)$ and $(1,1),$ among which only (1,1) could be dangerous. So we are allowed to neglect the other terms in a study of vector superluminality. The effective quartic Lagrangian then becomes
\begin{equation} \label{eq:3.8}
\mathcal{L}_{A,\text{eff},\text{quartic}}= -\frac{1}{4} \left( F^{\mu \nu}F_{\mu \nu} + g\Pi^{\mu \nu}\Pi^{ \alpha \beta} F_{\mu \alpha}F_{\nu \beta} \right),
\end{equation}
where $g$ is a numerical constant that replaces the corresponding $C_{pqr}$.

\subsection{Equations of Motion}
By (\ref{eq:3.8}), the equations of motion for $A_\alpha$ is
\begin{equation} \label{eq:3.9}
\partial_{\mu}F^{\mu \alpha}+g\partial_{\mu}(\Pi^{\mu \nu}\Pi^{\alpha \beta} F_{\nu \beta})=0.
\end{equation}
By construction, $\partial_{\mu}(\Pi^{\mu \nu}\Pi^{\alpha \beta})$ is proportional to the third derivatives of the scalar field, which have dubious physical meanings since in dRGT the equation of motion for $\phi$ is always second order by the properties of Galileon self-interactions \cite{Nicolis:2008in}. Therefore, without loss of generality, we set
\begin{equation}\label{eq:3.10}
\partial_{\mu}\partial_{\nu}\partial_{\alpha}\phi=0,
\end{equation}
upon which the matrix $\Pi^{\mu \nu} \equiv \partial^{\mu}\partial^{\nu}\phi$ is constant. This ensures the existence of plane wave solutions for (\ref{eq:3.9}). Then the effective equation of motion becomes
\begin{equation}\label{eq:3.11}
\eta^{\alpha \beta}\partial^{\nu}F_{\nu \beta}+g\Pi^{\mu \nu}\Pi^{\alpha \beta}\partial_{\mu}F_{\nu \beta}=0.
\end{equation}
Note that the second term, 
\begin{equation} \label{eq:3.12}
\sim\Pi^{\mu \nu}\Pi^{\alpha \beta}\partial_{\mu}F_{\nu \beta}=\Pi^{\mu \nu}\Pi^{\alpha \beta}\partial_{\mu}\partial_\nu A_\beta -\Pi^{\alpha \beta}\partial_{\beta}(\Pi^{\mu \nu}\partial_{\mu}A_\nu),
\end{equation}
cannot be entirely removed by a gauge choice in general. Therefore, the dangerous term does not vanish unless $g=0.$ The $g=0$ case then necessitates a discussion at higher orders, which we will cover in section 4. In (\ref{eq:3.12}), one may choose the gauge condition $\Pi^{\mu \nu}\partial_{\mu}A_\nu=0$ to remove the corresponding part, but then the divergence term from $\partial^{\nu}F_{\nu \beta}$ will be nonzero. In the following subsection, we still work in the Lorenz gauge $\partial^{\mu}A_{\mu}=0$ to enforce the well-known condition $k^{\mu}\epsilon_{\mu}(\mathbf{k},\sigma)=0$ for polarization vectors.

\subsection{Constructing Superluminal Solutions}
At a given point in spacetime, we can always perform a global Lorentz transformation to diagonalize $\Pi^{\mu \nu}$ because it is symmetric. For simplicity, we consider a static background solution for the scalar field, or that
\begin{equation}\label{eq:3.13}
\Pi^{\mu \nu}=\left( \begin{array}{cccc}
0 & 0 & 0 & 0 \\
0 & \rho & 0 & 0 \\
0 & 0 & \rho & 0 \\
0 & 0 & 0 & p \end{array} \right),
\end{equation}
where $\rho$ and $p$ are constants under our assumption in section 3.2. Here $p=-2\rho+\mathcal{O}(\frac{1}{\Lambda_3^3}),$ where the correction comes from the terms in the scalar equation of motion other than $\Box\phi$ \cite{deRham:2011pt,deRham:2010ik,Hinterbichler:2011tt}.\footnote{In particular, in a free theory of helicity-2 and -0 modes ($c_3=1/6,d_5=-1/48$), no scalar self-interaction exists and $p=-2\rho$ is exact.}  

To seek plane wave solutions, we also define the Fourier transform, $\tilde{A}_{\mu}(k)$, of the vector field by
\begin{equation}\label{eq:3.14}
A_{\mu}(x)\equiv \frac{1}{(2\pi)^4}\int d^4ke^{ik \cdot x}\tilde{A}_{\mu}(k),~ k^{\mu}\equiv(\omega, \mathbf{k}).
\end{equation}
Then the equation of motion (\ref{eq:3.11}) in Fourier space is equivalent to
\begin{equation}\label{eq:3.15}
\left( \begin{array}{cccc}
-\omega^2+\mathbf{k}^2 & 0 & 0 & 0 \\
0 & \begin{array}{c}-\omega^2+k_1^2 \\
+(1+K)k_2^2+(1+M)k_3^2
\end{array} & -Kk_{1}k_{2} & -Mk_{1}k_{3} \\
0 & -Kk_{1}k_{2} & \begin{array}{c}-\omega^2+(1+K)k_1^2 \\
+k_2^2+(1+M)k_3^2 \end{array} & -Mk_{2}k_{3} \\
0 & -Mk_{1}k_{3} & -Mk_{2}k_{3} & \begin{array}{c}-\omega^2+(1+M)k_1^2 \\
+(1+M)k_2^2+k_3^2 \end{array}
\end{array} \right)
\left( \begin{array}{c} \tilde{A}_{0} \\ \tilde{A}_{1} \\ \tilde{A}_{2}  \\ \tilde{A}_{3} \end{array} \right)=0,
\end{equation}
with
\begin{equation}\label{eq:3.16}
K\equiv g\rho^2
\end{equation}
and
\begin{equation}\label{eq:3.17}
M\equiv g\rho p.
\end{equation}
To obtain (\ref{eq:3.15}), we have used the gauge condition $\partial_{\mu}A^{\mu}=0$. In order for nontrivial solutions to exist for the homogeneous system described by (\ref{eq:3.15}), the matrix
\begin{equation}\label{eq:3.18}
\Upsilon\equiv \left( \begin{array}{cccc}
-\omega^2+\mathbf{k}^2 & 0 & 0 & 0 \\
0 & \begin{array}{c}-\omega^2+k_1^2 \\
+(1+K)k_2^2+(1+M)k_3^2
\end{array} & -Kk_{1}k_{2} & -Mk_{1}k_{3} \\
0 & -Kk_{1}k_{2} & \begin{array}{c}-\omega^2+(1+K)k_1^2 \\
+k_2^2+(1+M)k_3^2 \end{array} & -Mk_{2}k_{3} \\
0 & -Mk_{1}k_{3} & -Mk_{2}k_{3} & \begin{array}{c}-\omega^2+(1+M)k_1^2 \\
+(1+M)k_2^2+k_3^2 \end{array}
\end{array} \right)
\end{equation}
must have zero eigenvalues. In fact, its eigenvalues and eigenvectors are, with due correspondence:
\begin{equation}\label{eq:3.19}
\begin{array}{l}
\lambda_{1}=-\omega^2+(1+g\rho^2)(k_1^2+k_2^2)+(1+g\rho p)k_3^2, \\
\lambda_{2}=-\omega^2+\mathbf{k}^{2}(1+g\rho p),\\ 
\lambda_{3}=\lambda_4=-\omega^2+\mathbf{k}^{2};\\
\\
v_1^\mu=
\left( \begin{array}{c}
0 \\
-k_{2} \\
k_{1} \\
0 \end{array} \right),
v_2^\mu=
\left( \begin{array}{c}
0 \\
-k_1k_{3} \\
-k_2k_3 \\
k_{1}^2+k_2^2 \end{array} \right),
v_3^\mu=
\left( \begin{array}{c}
0 \\
k_{1} \\
k_{2} \\
k_{3} \end{array} \right),
v_4^\mu=
\left( \begin{array}{c}
1 \\
0 \\
0 \\
0 \end{array} \right).
\end{array}
\end{equation}
Based on these we can work out the polarization vectors and their dispersion relations. 

First, we consider $\lambda_3=\lambda_4=-\omega^2+\mathbf{k}^{2}=0$, giving the exactly luminal relation $\omega^2=\mathbf{k}^2$. In this case, any arbitrary linear combination $v=av_3+bv_4$ is also an eigenvector with the same dispersion relation. However, imposing the Lorenz gauge condition $k^\mu v_{\mu}=0$ sets $b=a{\mathbf{k}^2}/{\omega}=a\omega.$ This implies that $v^{\mu}=ak^{\mu}$, which is not normalizable given the dispersion relation. As a well-known fact, this mode, proportional to the momentum itself, is merely a mathematical artifact originating from the residual gauge freedom allowed by the Lorenz gauge. Thus, no nontrivial longitudinal mode can exist in the solution.

On the other hand, one can easily check that the Lorenz gauge conditions $k_{\mu}v_1^\mu=k_{\mu}v_2^\mu=0$ are satisfied, so we can normalize $v_1$ and $v_2$ to obtain the two transverse polarization vectors, respectively:
\begin{equation} \label{eq:3.20}
\begin{array}{l}
\epsilon^{\mu}(\mathbf{k},1)=\frac{1}{\sqrt{k_1^2+k_2^2}}\left( 0,-k_2,k_1,0 \right)\equiv(0,\hat{\mathbf{n}}(\mathbf{k},1)), \\
\epsilon^{\mu}(\mathbf{k},2)=\frac{1}{\sqrt{\mathbf{k}^2(k_1^2+k_2^2)}}\left( 0,-k_1k_3,-k_2k_3, k_1^2+k_2^2 \right)\equiv(0,\hat{\mathbf{n}}(\mathbf{k},2)).
\end{array}
\end{equation}
These polarization vectors fulfill the following criteria ($\sigma, \sigma '\in\{1,2\}$):
\begin{description}
\item[(i) Orthonormality:] $\epsilon_{\mu}(\mathbf{k},\sigma)\epsilon^{\mu *}(\mathbf{k},\sigma ')=\delta_{\sigma \sigma '};$
\item[(ii) Lorenz gauge condition:]$k_\mu \epsilon^{\mu}(\mathbf{k},\sigma)=0$;
\item[(iii) Completeness relation:]
\begin{equation*}
\sum_{\sigma}\epsilon^{\mu}(\mathbf{k},\sigma)\epsilon^{\nu}(\mathbf{k},\sigma)^*=\left(
\begin{array}{cccc}
0 & 0 & 0 & 0 \\
0 & 1-\frac{k_1^2}{\mathbf{k}^2} & -\frac{k_1k_2}{\mathbf{k}^2} & -\frac{k_1k_3}{\mathbf{k}^2}\\
0 & -\frac{k_1k_2}{\mathbf{k}^2} & 1-\frac{k_2^2}{\mathbf{k}^2}  & -\frac{k_2k_3}{\mathbf{k}^2}\\
0 & -\frac{k_1k_3}{\mathbf{k}^2} & -\frac{k_2k_3}{\mathbf{k}^2} &1-\frac{k_3^2}{\mathbf{k}^2} 
\end{array} \right),
\end{equation*}
i.e. $\sum_{\sigma}\hat{\text{n}}_i(\mathbf{k},\sigma)\hat{\text{n}}_j^*(\mathbf{k},\sigma)=\delta_{ij}-\frac{k_ik_j}{\mathbf{k}^2}$.
\end{description}
So the solution is consistent with the physics of massless vector fields. The fact that only two modes exist can be seen more generally in a degree of freedom count, which we will present in section 4.

Now we are at a point to reveal superluminality for the transverse modes. We have the following two possibilities:
\begin{description}
  \item[Case I:]$\sigma=1$. In order for this mode to propagate, we impose $\lambda_{1}=0.$ The dispersion relation is
  \begin{align} \label{eq:3.21}
  \omega^2 &=  (1+g\rho^2)(k_1^2+k_2^2)+(1+g\rho p)k_3^2 \nonumber \\
  &\approx(1+g\rho^2)(k_1^2+k_2^2)+(1-2g\rho^2)k_3^2.
  \end{align}
  
If $g>0$, then in the case $k_1^2+k_2^2 \gg k_3^2$, $\omega^2 \approx (1+g\rho^2)(k_1^2+k_2^2)\approx (1+g\rho^2)\mathbf{k}^2$ is superluminal. If $g<0$, then in the case $k_3^2 \gg k_1^2+k_2^2$, $\omega^2 \approx (1-2g\rho^2)k_3^2\approx (1-2g\rho^2)\mathbf{k}^2$ is superluminal.
  \item[Case II:] $\sigma=2$. In order for this mode to propagate, we impose $\lambda_{2}=0.$ Then $\omega^2=(1+g\rho p)\mathbf{k}^2\approx (1-2g\rho^2)\mathbf{k}^2,$ which can be superluminal if $g<0$.
\end{description}
Therefore, up to the quartic order in the fields, superluminal vector excitation modes exist if and only if $g \neq 0,$ and the conditions by which a mode or modes exhibit superluminality are determined by the sign of $g$.

Since a dRGT theory is specified by the values of free parameters $c_3$ and $d_5$ (or equivalently, $\alpha_3$ and $\alpha_4$ in the language of Refs. \cite{Hassan:2011hr,Gabadadze:2013ria,Ondo:2013wka}, with slightly different definitions), it suffices to figure out the algebraic dependence of $g$ on the free parameters. This can be done by extracting the term $\sim \Pi^{\mu \nu}\Pi^{\alpha \beta}F_{\mu \alpha}F_{\nu \beta}=-[\Pi F \Pi F]$ from Eqn. (3.36) of Ref. \cite{Ondo:2013wka}. For example, in the minimal model specified by $c_3=1/6$ and $d_5=-1/48$, we find that $g=\frac{1}{2\Lambda_3^6} >0$, implying that the $\sigma=1$ mode described above exhibits superluminality if the vector's momentum has sufficiently small 3-component.

In addition, and for the rigor of the argument, note that 
\begin{equation} \label{eq:3.180}
k^{\mu}=(\omega,0,0,k_3)
\end{equation}
are the singular points in the momentum space for the polarization vectors in (\ref{eq:3.20}). So we need to consider this situation in a separate case. For such momenta, (\ref{eq:3.18}) is reduced to
\begin{equation}\label{eq:3.181}
\Upsilon\equiv \left( \begin{array}{cccc}
-\omega^2+k_3^2 & 0 & 0 & 0 \\
0 & -\omega^2+(1+M)k_3^2 & 0 & 0 \\
0 & 0 & -\omega^2+(1+M)k_3^2 & 0 \\
0 & 0 & 0 & -\omega^2+k_3^2
\end{array} \right),
\end{equation}
whose eigenvalues and eigenvectors are, with due correspondence:
\begin{equation} \label{eq:3.182}
\begin{array}{l}
\lambda_{1}=\lambda_2=-\omega^2+k_3^{2}(1+g\rho p),\\ 
\lambda_{3}=\lambda_4=-\omega^2+k_3^{2};\\
\\
v_1^\mu=
\left( \begin{array}{c}
0 \\
1 \\
0 \\
0 \end{array} \right),
v_2^\mu=
\left( \begin{array}{c}
0 \\
0 \\
1 \\
0 \end{array} \right),
v_3^\mu=
\left( \begin{array}{c}
0 \\
0 \\
0 \\
1 \end{array} \right),
v_4^\mu=
\left( \begin{array}{c}
1 \\
0 \\
0 \\
0 \end{array} \right).
\end{array}
\end{equation}
In the case $\lambda_{3}=\lambda_4=-\omega^2+k_3^{2}=0,$ it is easy to see that any linear combination of $v_3$ and $v_4$ which also fulfills the Lorenz gauge condition must be proportional to $(1,0,0,1),$ which is just a gauge mode following from the residual gauge freedom of the Lorenz gauge. Thus no nontrivial longitudinal mode can exist. For $\lambda_{1}=\lambda_2=-\omega^2+k_3^{2}(1+g\rho p)=0,$ it is natural to take the polarization vectors as
\begin{equation}\label{eq:3.183}
\begin{array}{l}
\epsilon^{\mu}(\mathbf{k},1)=(0,1,0,0),\\
\epsilon^{\mu}(\mathbf{k},2)=(0,0,1,0).
\end{array}
\end{equation}
It is straightforward to check that they satisfy the orthonormality, Lorenz gauge, and completeness relations. They share a common dispersion relation $\omega^2=(1+g\rho p)k_3^{2}=(1+g\rho p)\mathbf{k}^2\approx (1-2g\rho^2)\mathbf{k}^2$, which can be superluminal if $g<0$. In the minimal model, $g=\frac{1}{2\Lambda_3^6}>0,$ and these transverse modes are safely subluminal.

The scaling (\ref{eq:2.10}) shows that $g$ is always suppressed by the sixth power of the scale $\Lambda_3$. This implies that the superluminal shift in the propagation speed is of the order
\begin{equation}\label{eq:3.26}
\Delta c \sim \frac{\rho^2}{\Lambda_{3}^6}.
\end{equation}
A construction of superluminal solutions in a homogeneous scalar background can be found in Appendix A.
   
\section{Vector Superluminality to All Orders}
\subsection{Leading Interaction and Degree of Freedom Count}

In this section, our goal is to obtain generalizations of the results derived in section 3 to all orders. Given the formally resummed vector Lagrangian specified by (\ref{eq:3.1}) and (\ref{eq:3.4}) and the line of argument leading to (\ref{eq:3.7}), we see that in general, all that matters to the issue of vector superluminality are the leading nonzero interaction terms with $p\geq 1$.\footnote{Of course, in which case all dangerous terms at lower orders are set to zero by the choice of free parameters, if possible.} For convenience, we define $s$ as the order in the fields of such terms, so that our analysis in section 3 covers the case $s=4$.\footnote{Due to the limited availability of free parameters (only two in the dRGT massive gravity), there might be an upper limit for the value of $s$.} Note that there could be more than one such terms in a model with fixed free parameters. For instance, at the hexic order in the fields, both the terms with $(p,q)=(1,3)$ and $(2,2)$ can contribute to superluminality. Meanwhile, there are only finitely many such terms for a given $s$, since the set of allowed pairings $\{ (p,q)|s \}$, of which the dangerous pairings are elements, is finite. Thus, the generalized effective Lagrangian is
\begin{align} \label{eq:4.1}
\mathcal{L}_{A,\text{eff}} &= -\frac{1}{4} \left( F^{\mu \nu}F_{\mu \nu} +\sum_{n=1}^N g_n F_n(\Pi)(\Pi^{\mu \nu})^{p_n}(\Pi^{ \alpha \beta})^{q_n} F_{\mu \alpha}F_{\nu \beta} \right) \nonumber \\
&\equiv  -\frac{1}{4} \left( F^{\mu \nu}F_{\mu \nu} + G^{\mu \nu \alpha \beta} F_{\mu \alpha}F_{\nu \beta} \right),
\end{align}
where the relabeled indices $\{ n \}$ run over all dangerous terms and $g_n \neq 0$ for all $n$.

After this, we can, of course, repeat what we have done in section 3: derive equations of motion, choose a representative background scalar solution and work out the superluminal modes. However, this procedure can be very cumbersome given the complicated form of (\ref{eq:4.1}) and only be carried out after choosing a gauge (In section 3, it is the Lorenz gauge). Instead, we proceed with casting the effective action into Hamiltonian form and counting the number of degrees of freedom. We will show that it propagates 2 degrees of freedom, the right number for a massless vector.

To begin with, we can perform a global Lorentz transformation to diagonalize the symmetric tensor $\Pi^{\mu \nu}$ at a given spacetime point. Then we Legendre transform (\ref{eq:4.1}) only with respect to the spatial components $A_i$. The canonical momenta are
\begin{align}\label{eq:4.2}
\pi_i=\frac{\partial \mathcal{L}_{A,\text{eff}}}{\partial \dot{A}_i}&=F_{0i}-G^{0\nu i \beta}F_{\nu \beta} \nonumber \\
&=(1-G^{00ii})F_{0i} \nonumber \\
&=(1-G^{00ii})(\dot{A}_i-\partial_iA_0),
\end{align}
where in the last two lines $i$'s are not summed. Inverting, we have
\begin{equation} \label{eq:4.3}
\dot{A}_i=\frac{\pi_i}{1-G^{00ii}}+\partial_iA_0.
\end{equation}
In terms of these Hamiltonian variables, (\ref{eq:4.1}) becomes
\begin{equation} \label{eq:4.4}
\mathcal{L}_{A,\text{eff}} = -\frac{1}{4} \left[\sum_{i,j}(1+G^{iijj})F_{ij}F_{ij}+2\sum_{k}\left( 1+G^{00kk} \right) \left( \frac{\pi_{k}}{1-G^{00kk}} \right)^2 \right]\equiv -\mathcal{H}_1,
\end{equation}
which is independent of $A_0.$ Then the generalized effective action becomes
\begin{align}\label{eq:4.5}
S_{A,\text{eff}}&=\int d^4x \left[ \pi_i\dot{A}_i-\mathcal{H}_1-\sum_{k}\pi_k\left(\frac{\pi_k}{1-G^{00kk}}+\partial_kA_0\right) \right] \nonumber \\
&\xrightarrow{\text{by~parts}}\int d^4x \left[ \pi_i\dot{A}_i-\left( \underbrace{\mathcal{H}_1+\sum_{k} \frac{\pi_k^2}{1-G^{00kk}}}_{\equiv\mathcal{H}} \right) +A_0 \partial_i \pi_i \right].
\end{align}
Since the timelike component $A_0$ is multiplied by terms with no time derivatives, we can regard it as a Lagrange multiplier enforcing the single constraint $\partial_iA_i=0,$ which is the familiar Coulomb gauge. Clearly, this is a first class constraint, and the Hamiltonian $\mathcal{H}$ defined by (\ref{eq:4.5}) is first class. So the action represents a first class gauge system. The $A_i$ and $\pi_i$ have three components each, so they span a 6-dimensional phase space. The constraint $\partial_iA_i=0$ then yields a 5-dimensional constraint surface. Also, the constraint generates a gauge invariance, giving 1-dimensional gauge orbits, so that the gauge invariant quotient by the orbits is 4-dimensional.\footnote{An introduction to the terminology used here can be found in Ref. \cite{Henn}.} These are the two polarizations of the massless vector along with their conjugate momenta. This confirms the correctness of our explicit mode solutions obtained in the special case $s=4$.

\subsection{Is the Superluminality Physical?}
Now we have the knowledge that the effective action represents a first class gauge system and gives rise to two transverse modes. This enables us to use power count in testing physicality for the superluminal modes. The structure of (\ref{eq:4.1}) shows that any superluminal mode, if existent as a solution to the effective equations of motion, has a shift in the propagation speed of the order
\begin{equation}\label{eq:4.6}
\Delta c \sim \frac{\rho^{s-2}}{\Lambda_3^{3s-6}},
\end{equation}
where $\rho$ is a typical value of the magnitude of $\Pi^{\mu \nu}.$\footnote{The argument presented in this section applies for all $s \ge 3,$ so it covers the gauge-dependent results in sections 2 and 3 as well.}  In particular, (\ref{eq:3.26}) gives the special case  $s=4$.
 
For this result to be consistent, we have to check that the superluminal effect caused by the shift in propagation speed is physically detectable. This dictates that the gain of the superluminal propagating mode over an exactly luminal signal in its course of traveling the typical distance by which the background vector solution varies must be at least of order of one wavelength \cite{Nicolis:2009qm}. In short, we need
\begin{equation} \label{eq:4.7}
\Delta c ~k L' \gtrsim 1.
\end{equation}
Here we denote the length scale of variation for vector mode by $L'$, and in the following discussion, that for scalar mode $L$. Since both the scalar and the vector modes are the St\"ukelberg fields of a massive graviton, we expect them to have similar length scale of variation, $L' \sim L$. Plugging (\ref{eq:4.6}) into the l.h.s of (\ref{eq:4.7}), we have 
\begin{equation} \label{eq:4.8}
\Delta c ~k L' \sim 
\frac{(\partial^2\phi)^{s-2}}{\Lambda_{3}^{3s-6}} \cdot k L' \sim \frac{\phi^{s-2}L^{4-2s}}{\Lambda_{3}^{3s-6}} \cdot k L'.
\end{equation}

Note that the generalized effective Lagrangian in (\ref{eq:4.1}) (and thus the quartic order case discussed in section 3) is obtained from a classical perturbative method based on order-by-order calculation, so we first check whether the superluminal signal is observable within the region where the perturbative expansion is valid. Therefore, locally we demand that 
\begin{equation} \label{eq:4.9}
\phi \ll \Lambda_3,~ L\gg 1/\Lambda_3.
\end{equation}
These conditions then imply that
\begin{equation} \label{eq:4.10}
\Delta c ~k L' \sim 
 \frac{\phi^{s-2}(L^{-1})^{2s-4}}{\Lambda_{3}^{3s-6}} \cdot k L
\ll \frac{\Lambda_3^{s-2} \cdot \Lambda_3^{2s-5} \cdot k}{\Lambda_{3}^{3s-6}}=\frac{k}{\Lambda_3},
\end{equation}
from which it is clear that if the vector momentum remains in the linear region of the perturbation theory ($k \ll \Lambda_3$), the superluminal signal would not be detectable.

It may seem that a large vector momentum does not break the perturbativity of the effective Lagrangian (\ref{eq:4.1}), where the vector excitation modes only show up quadratically. However, higher order terms in the vector field would appear when one moves out of the decoupling limit, and thus for concerns of continuity, the case of large $k$ necessitates a discussion in the strong coupling region.

It then follows that all physically observable superluminal propagating signals, if any, must have frequencies at least comparable to the strong coupling scale of the dRGT massive gravity. In the strong coupling region, while the perturbation theory breaks down, the classical estimate made in (\ref{eq:4.6}) can still be trusted since no new operators enter the physical picture, only that the non-linear operators become important \cite{Nicolis:2004qq,Hinterbichler:2011tt,deRham:2014zqa}. This change can be addressed by a redefinition of $s$, which essentially characterizes the most important operators affecting superluminal solutions, and these operators are already included  in the effective Lagrangian (\ref{eq:4.1}). In the non-linear regime, we have at least
\begin{equation} \label{eq:4.11}
\phi \sim \Lambda_3,~ L,L' \sim 1/\Lambda_3.
\end{equation}  
This implies that
\begin{equation} \label{eq:4.12}
\Delta c ~k L' \sim 
 \frac{\phi^{s-2}(L^{-1})^{2s-4}}{\Lambda_{3}^{3s-6}} \cdot k L
\sim \frac{\Lambda_3^{s-2} \cdot \Lambda_3^{2s-5} \cdot k}{\Lambda_{3}^{3s-6}}=\frac{k}{\Lambda_3}
\gtrsim 1
\end{equation}
as long as the magnitude of $k$ lies in the strong coupling region. As such, the vector superluminality characterized by (\ref{eq:4.6}) can be physically consistent in the classical non-linear regime of the effective field theory.

Furthermore, due to the Vainshtein mechanism in dRGT, for background field configurations satisfying $\partial^2\phi \gg \Lambda_3^3$, the strong coupling gets rescaled and assumes a much larger value that could be several orders higher than $\Lambda_3$. In the past, the redressing mechanism of the strong coupling scale has been demonstrated in the helicity-0 Galileon models \cite{Nicolis:2008in,deRham:2014zqa}. In the scalar sector, the new strong coupling scale is symbolically $\Lambda_{\star} \sim Z^{1/2}\Lambda_3,$ where $Z^{\mu \nu}(\pi_0)$ is the effective metric for scalar fluctuations around a background solution $\pi_0$. Similarly, the redressed strong coupling scale also manifests itself through the vector Lagrangian in the form of (\ref{eq:3.1}), where the tensor $\mathcal{T}^{\mu \nu \alpha \beta}$ goes like powers of $\frac{\partial^2\phi}{\Lambda_3^3}$ and takes large values in the strong coupling region. To see this mechanism for vector mode more explicitly, we go a little outside the decoupling limit, so the vector coupling to a source does not get eliminated. Plugging (\ref{eq:7}) and (\ref{eq:2.10}) into the original massive spin-2 source term $\frac{1}{M_{\text{p}}}H_{\mu \nu}T^{\mu \nu}$, the Lagrangian for vector fluctuation becomes
\begin{equation}
\mathcal{L}_{A}=-\frac{1}{4} \mathcal{T}^{\mu \nu \alpha \beta} F_{\mu \alpha}F_{\nu \beta}
-\frac{\partial_{\mu}A^{\alpha}\partial_{\nu}A_{\alpha}}{M_{\text{p}}\Lambda_3^3} \frac{\delta T^{\mu \nu}}{M_{\text{p}}},
\end{equation}
where $\delta T^{\mu \nu}$ is a perturbation giving rise to vector fluctuations. Then symbolically, performing a canonical normalization for the vector mode results in an increase of the coupling scale by the factor of $\mathcal{T}^{1/2}$.\footnote{Note that since we are outside the decoupling limit here, strictly speaking, operators more than quadratic in the vector field could contribute to $\mathcal{T}^{\mu \nu \alpha \beta}$, yet the mechanism by which the coupling scale gets redressed remains valid.}
Thus the vector mode frequencies $k \sim \Lambda_3$ are safely below the redressed strong coupling scale, reinforcing the point that the vector superluminality obtained previously can be trusted at the classical level. 

\section{Discussion and Outlook}
For the scalar fields in dRGT, it has been shown that the cubic and higher order Galileon terms inevitably lead to superluminal group velocities  \cite{Nicolis:2008in,Nicolis:2009qm,deFromont:2013iwa,Adams:2006sv}, and one needs further considerations at the level of asymptotic conditions \cite{deFromont:2013iwa} or closed timelike curves (CTCs) \cite{Burrage:2011cr} to explore whether the scalar superluminalities are physically consistent. For the vector fields, we see that while the superluminal fluctuations in the perturbation theory do not produce measurable propagating signals, they could be a physical possibility in the strong coupling region of dRGT where non-linear operators dominate. Fundamentally, this is in agreement with the well-received observation that the non-linearities in interacting massive gravity bring about superluminality, as both GR and FP theories are free of tachyonic propagating modes.

Still, the issue of superluminality remains inconclusive for both scalar and vector fields in the full dRGT theory. To determine whether superluminal propagating speeds eventually lead to causality violation, the ultimate measure is the high frequency limit of phase velocity, $v_{\text{ph}}(\infty)$, which is also called the front velocity \cite{Shore:2007um,deRham:2014zqa}. In order to compute the front velocity, one must work in an energy range even beyond the strong coupling region. This is a regime where quantum corrections would dominate over the classical operators. Therefore, the vector superluminalities computed previously, which are only low frequency group and phase velocities, would no longer be valid for considerations at the quantum level. The task of testifying the (a)causality of dRGT then requires a full knowledge of the UV completion scheme for the effective field theory.

Besides, it has been claimed that if the Lagrangian of a field involves only self-interactions, the existence of superluminal propagating modes is exclusively determined by the leading order interaction term \cite{Adams:2006sv}. Indeed, for the scalar mode in the decoupling limit dGRT, this statement is sufficient for the purpose of studying superluminality, since the absence of terms linear in vector field allows one to safely set the vectors to zero at the classical level. But we cannot do the same thing when studying vector superluminality because of the tadpole cancellation condition for the scalar field. As vector self-interaction terms are absent in the decoupling limit, the analysis in this paper seems to suggest that we need to modify the statement above: vector superluminalities are determined not by the leading mixing terms (the cubic terms, which are safe), but the relevant quartic or higher order terms in the form $\sim[\Pi F \Pi F]$, $[\Pi F \Pi F \Pi]$, etc.\footnote{Terms with two $F$'s contracted together, like $\sim[\Pi F F \Pi]$ and $\sim[\Pi F F \Pi \Pi]$, do not affect vector superluminality since they contain a factor of $\eta^{\mu \nu}$.} If we go outside the decoupling limit, vector self-interaction terms like $\sim(F^{\mu \nu}F_{\mu \nu})^2$,$(F^{\mu \nu}\tilde{F}_{\mu \nu})^2$ will reappear, and the previous statement in Ref. \cite{Adams:2006sv} about the relations between superluminality and leading order terms are true again. It could be a good practice to explore vector superluminalities beyond the decoupling limit by collecting these terms from the St\"{u}kelberg expansion of dRGT Lagrangian. It could also be interesting to look at the issue of vector superluminality in other cosmological models of gravity, like the partially massless gravity in de Sitter background, described in detail in Refs. \cite{deRham:2012kf,Deser:2013uy,deRham:2013wv}.\footnote{See Refs. \cite{Koyama:2011wx,Tasinato:2012ze} for relevant results about the (A)dS vector Lagrangian in the decoupling limit.}

\acknowledgments
The author would like to thank Alberto Nicolis, Rachel A. Rosen and Claudia de Rham for valuable discussions, as well as Allan Blaer for support and encouragement. The author has also benefited from communications with Andrew Waldron on topics related to the present work. The author is funded in part by the Columbia Undergraduate Scholars Program.

\appendix
\section{Quartic Order Superluminal Solutions in a Different Background}
Here we assume a homogeneous background solution for the scalar field, or that
\begin{equation} \label{eq:A1}
\Pi^{\mu \nu}=\left( \begin{array}{cccc}
p & 0 & 0 & 0 \\
0 & \rho & 0 & 0 \\
0 & 0 & \rho & 0 \\
0 & 0 & 0 & \rho \end{array} \right),
\end{equation}
where $p$ and $\rho$ are constants and $p=3\rho+\mathcal{O}(\frac{1}{\Lambda_3^3}).$ In the minimal model, $p=3\rho$ is exact.

Plugging (\ref{eq:A1}) into (\ref{eq:3.11}) and transforming into Fourier space, we get
\begin{equation} \label{eq:A2}
\left( \begin{array}{cccc}
\omega^2-\mathbf{k}^2 & 0 & 0 & 0 \\
0 & K+Mk_{1}^2 & Mk_{1}k_{2} & Mk_{1}k_{3} \\
0 & Mk_{1}k_{2} & K+Mk_{2}^2 & Mk_{2}k_{3} \\
0 & Mk_{1}k_{3} & Mk_{2}k_{3} & K+Mk_{3}^2 \end{array} \right)
\left( \begin{array}{c}  \tilde{A}_{0} \\ \tilde{A}_{1} \\ \tilde{A}_{2}  \\ \tilde{A}_{3} \end{array} \right)=0,
\end{equation}
with
\begin{equation}
K\equiv (1-gp\rho)\omega^2-(1+g\rho^2)\mathbf{k}^{2}
\end{equation}
and
\begin{equation}
M\equiv g\rho(p+\rho),
\end{equation}
where we have used the gauge condition $\partial_{\mu}A^{\mu}=0$. In order for nontrivial solutions to exist for the homogeneous system described by (\ref{eq:A2}), the matrix
\begin{equation}\label{eq:A5}
\Upsilon\equiv \left( \begin{array}{cccc}
\omega^2-\mathbf{k}^2 & 0 & 0 & 0 \\
0 & K+Mk_{1}^2 & Mk_{1}k_{2} & Mk_{1}k_{3} \\
0 & Mk_{1}k_{2} & K+Mk_{2}^2 & Mk_{2}k_{3} \\
0 & Mk_{1}k_{3} & Mk_{2}k_{3} & K+Mk_{3}^2 \end{array} \right)
\end{equation}
must have zero eigenvalues. Its eigenvalues and eigenvectors are, with due correspondence:
\begin{equation}\label{eq:A6}
\begin{array}{l}
\lambda_{1}=\lambda_{2}=(1-g\rho p)\omega^2-(1+g\rho^2)\mathbf{k}^{2},\\ 
\lambda_{3}=(1-g\rho p)(\omega^2-\mathbf{k}^{2}), \\
\lambda_4=\omega^2-\mathbf{k}^{2};\\
\\
v_1^\mu=
\left( \begin{array}{c}
0 \\
-k_{2} \\
k_1 \\
0 \end{array} \right),
v_2^\mu=
\left( \begin{array}{c}
0 \\
-k_{3} \\
0\\
k_{1} \end{array} \right),
v_3^\mu=
\left( \begin{array}{c}
0 \\
k_{1} \\
k_{2} \\
k_{3} \end{array} \right),
v_4^\mu=
\left( \begin{array}{c}
1 \\
0 \\
0 \\
0 \end{array} \right).
\end{array}
\end{equation}
Based on these we can work out the polarization vectors and their dispersion relations. 

First, we see that $v_3$ and $v_4$ share the same dispersion relation $\omega^2=\mathbf{k}^{2}$. In this case, any arbitrary linear combination $v=av_3+bv_4$ is also an eigenvector with the same dispersion relation, because the dispersion relation sets the eigenvalues to zero. Then we impose the Lorenz gauge condition $k^\mu v_{\mu}=0$, which sets $b=a{\mathbf{k}^2}/{\omega}=a\omega.$ This implies that $v^{\mu}=ak^{\mu}$, which represents merely a gauge mode. As such, no nontrivial longitudinal mode can exist.

On the other hand, one can easily check that the Lorenz gauge conditions $k_{\mu}v_1^\mu=k_{\mu}v_2^\mu=0$ are satisfied, so we can normalize $v_1$ and $v_2$ to obtain the two transverse polarization vectors, respectively:
\begin{equation} \label{eq:A7}
\begin{array}{l}
\epsilon^{\mu}(\mathbf{k},1)=\frac{1}{\sqrt{k_1^2+k_2^2}}\left( 0,-k_2,k_1,0 \right)\equiv(0,\hat{\mathbf{n}}(\mathbf{k},1)), \\
\epsilon^{\mu}(\mathbf{k},2)=\frac{1}{\sqrt{k_1^2+k_3^2}}\left( 0,-k_3,0,k_1 \right)\equiv(0,\hat{\mathbf{n}}(\mathbf{k},2)).
\end{array}
\end{equation}
Similar to section 3.3, these polarization vectors fulfill the following criteria ($\sigma, \sigma '\in\{1,2\}$):
\begin{description}
\item[(i) Orthonormality:] $\epsilon_{\mu}(\mathbf{k},\sigma)\epsilon^{\mu *}(\mathbf{k},\sigma ')=\delta_{\sigma \sigma '};$
\item[(ii) Lorenz gauge condition:]$k_\mu \epsilon^{\mu}(\mathbf{k},\sigma)=0$;
\item[(iii) Completeness relation:]
\begin{equation*}
\sum_{\sigma}\epsilon^{\mu}(\mathbf{k},\sigma)\epsilon^{\nu}(\mathbf{k},\sigma)^*=\left(
\begin{array}{cccc}
0 & 0 & 0 & 0 \\
0 & 1-\frac{k_1^2}{\mathbf{k}^2} & -\frac{k_1k_2}{\mathbf{k}^2} & -\frac{k_1k_3}{\mathbf{k}^2}\\
0 & -\frac{k_1k_2}{\mathbf{k}^2} & 1-\frac{k_2^2}{\mathbf{k}^2}  & -\frac{k_2k_3}{\mathbf{k}^2}\\
0 & -\frac{k_1k_3}{\mathbf{k}^2} & -\frac{k_2k_3}{\mathbf{k}^2} &1-\frac{k_3^2}{\mathbf{k}^2} 
\end{array} \right),
\end{equation*}
i.e. $\sum_{\sigma}\hat{\text{n}}_i(\mathbf{k},\sigma)\hat{\text{n}}_j^*(\mathbf{k},\sigma)=\delta_{ij}-\frac{k_ik_j}{\mathbf{k}^2}$.
\end{description}
So they are consistent with the physics of massless vectors.

Next, by setting $\lambda_1=\lambda_2=0$, we find that the two transverse modes share the same dispersion relation
  \begin{align}
  \omega^2 &=\frac{1+g\rho^2}{1-g\rho p}\mathbf{k}^{2} \nonumber \\
  &=(1+g\rho(p+\rho))\mathbf{k}^{2}+\mathcal{O}(g^2) \nonumber \\
  &\approx (1+4g\rho^2)\mathbf{k}^{2}.
  \end{align}
which can be superluminal if $g>0$. In the minimal model, $g=\frac{1}{2\Lambda_3^6}>0$, so both modes exhibit superluminality.

For completeness, we see that $k^{\mu}=(\omega,0,0,k_3)$ or $k^{\mu}=(\omega,0,k_2,0)$ makes a polarization vector in (\ref{eq:A7}) singular. Since the background scalar solution is homogeneous, all three spatial directions are on the equal footing (this can also be seen from (\ref{eq:A5})). Without loss of generality, it suffices to consider $k^{\mu}=(\omega,0,0,k_3)$ here, under which (\ref{eq:A5}) becomes
\begin{equation} \label{eq:A9}
\Upsilon\equiv \left( \begin{array}{cccc}
\omega^2-k_3^2 & 0 & 0 & 0 \\
0 & J & 0 & 0 \\
0 & 0 & J & 0 \\
0 & 0 & 0 & J+Mk_{3}^2 \end{array} \right),
\end{equation}
with $J\equiv(1-gp\rho)\omega^2-(1+g\rho^2){k}_3^{2}.$
The eigenvalues and eigenvectors are, with due correspondence:
\begin{equation}\label{eq:A10}
\begin{array}{l}
\lambda_{1}=\lambda_{2}=(1-g\rho p)\omega^2-(1+g\rho^2)k_3^{2},\\ 
\lambda_{3}=(1-g\rho p)(\omega^2-{k}_3^{2}), \\
\lambda_4=\omega^2-{k}_3^{2};\\
\\
v_1^\mu=
\left( \begin{array}{c}
0 \\
1 \\
0 \\
0 \end{array} \right),
v_2^\mu=
\left( \begin{array}{c}
0 \\
0 \\
1\\
0 \end{array} \right),
v_3^\mu=
\left( \begin{array}{c}
0 \\
0 \\
0 \\
1 \end{array} \right),
v_4^\mu=
\left( \begin{array}{c}
1 \\
0 \\
0 \\
0 \end{array} \right).
\end{array}
\end{equation}
Here $v_3$ and $v_4$ have the same dispersion relation $\omega^2=k_3^2.$
In this case, it is easy to see that any linear combination of $v_3$ and $v_4$ which also fulfills the Lorenz gauge condition must be proportional to $(1,0,0,1),$ which is a gauge mode. Hence no nontrivial longitudinal mode exists. For $\lambda_{1}=\lambda_2=(1-g\rho p)\omega^2-(1+g\rho^2)k_3^{2}=0,$ it is natural to take the polarization vectors as
\begin{equation}\label{eq:A11}
\begin{array}{l}
\epsilon^{\mu}(\mathbf{k},1)=(0,1,0,0),\\
\epsilon^{\mu}(\mathbf{k},2)=(0,0,1,0).
\end{array}
\end{equation}
It is straightforward to check that they satisfy the orthonormality, Lorenz gauge, and completeness relations. They share a common dispersion relation $\omega^2 =\frac{1+g\rho^2}{1-g\rho p}{k}_3^{2} 
 =(1+g\rho(p+\rho)){k}_3^{2}+\mathcal{O}(g^2)\approx (1+4g\rho^2){k}_3^{2}$, which can be superluminal if $g>0$. In the minimal model, $g=\frac{1}{2\Lambda_3^6}>0,$ so both transverse modes are superluminal.

The discussion of the physicality of these superluminal modes follows that presented in section 4.2.

\end{document}